# Tracking single atoms in a liquid environment

Nick Clark[1,3], Daniel J. Kelly[1,3], Mingwei Zhou[2,3], Yi-Chao Zou[1,3,4], Chang Woo Myung[5], David G. Hopkinson[1,3], Christoph Schran[5], Angelos Michaelides[5], Roman Gorbachev[2,3]* and Sarah J. Haigh[1,3^]

[1]*Department of Materials*, [2]*Department of Physics and Astronomy*, and [3]*National Graphene Institute, University of Manchester, Oxford Road, Manchester M13 9PL, United Kingdom.* [4]*School of Materials Science and Engineering, Sun Yat-sen University, Guangzhou, 510275, PR China.* [5]*Yusuf Hamied Department of Chemistry, University of Cambridge, Lensfield Road, Cambridge CB2 1EW, United Kingdom.*

*roman@manchester.ac.uk

^sarah.haigh@manchester.ac.uk

**The chemical behaviour of single metal atoms largely depends on the local coordination environment, including interactions with the substrate and with the surrounding gas or liquid. However, the key instrumentation for studying such systems at the atomic scale generally requires high vacuum conditions, limiting the degree to which the aforementioned environmental parameters can be investigated. Here we develop a new platform for transmission electron microscopy investigation of single metal atoms in liquids and study the dynamic behaviour of individual platinum atoms on the surface of a single layer MoS$_2$ crystal in water. To achieve the record single atom resolution, we introduce a double liquid cell based on a 2D material heterostructure, which allows us to submerge an atomically thin membrane with liquid on both sides while maintaining the total specimen thickness of only ~ 70 nm. By comparison with an identical specimen imaged under high vacuum conditions, we reveal drastic differences in the single atom resting sites and atomic hopping behaviour, demonstrating that *in situ* imaging conditions are essential to gain complete understanding of the chemical activity of individual atoms. These findings pave the way for *in situ* liquid imaging of chemical processes with single atom precision.**

Single atom catalysis is one of the most active and exciting frontiers in heterogeneous catalysis[1-3]. Atomically dispersed metal catalysts exhibit exceptional activity and selectivity for a wide range of reactions with minimal precious metal content compared to conventional catalysts[4,5]. Rapid advancement of this field has been aided by recent developments in characterisation and modelling techniques, which have guided new synthetic pathways for high performing catalysts, as well more efficient and cost-effective production methods[6,7]. Understanding the behaviour of single atoms in liquids is also key to the modelling the early stages of nucleation[8] and many other chemical processes. High angle annular dark-field scanning transmission electron microscopy (HAADF STEM) has emerged as one of the most valuable characterisation techniques for such studies due to its ability to directly visualise single metal atoms on a variety of substrates[9,10]. Although most transmission electron microscopy (TEM) is performed under high vacuum conditions, the last decade has seen major developments in *in situ* TEM; where specimens are imaged in liquid or gaseous environments that mimic reaction conditions[11,12]. The importance of this approach is exemplified by key studies demonstrating that catalytic nanoparticles have different surface structures in high vacuum than when exposed to a reaction gas[13,14], and that local coordination can evolve with applied environmental conditions, directly impacting catalytic performance[15].

Liquid cell *in situ* TEM is a means of imaging dynamic processes in liquid environments by encapsulating the specimen, along with a thin layer of liquid, between two electron transparent windows[11,16]. The imaging resolution for commercial liquid cells is limited to a few nanometres due to electron scattering in the silicon nitride (SiN$_x$) windows and in the liquid layer[17]. The development of graphene liquid cells, on the other hand, has enabled atomic resolution imaging of metal nanoparticles in liquids[18], showing that graphene is an ideal window material due to its extreme thinness, high mechanical strength[19], low atomic number, chemical inertness, impermeability[20], and its ability to scavenge aggressive radical species[21,22]. However, initial graphene liquid cell (GLC) designs[23] relied on serendipitous formation of liquid pockets between two graphene sheets, and exhibited low yield and poor stability under prolonged electron exposure[24]. More advanced designs have incorporated a patterned spacer layer of $SiN_x$ or hexagonal boron nitride (hBN) to define liquid pockets[25,26] providing improved control of the GLC geometry that allow for more control over experimental conditions. Nonetheless, the ability to image individual atoms in liquids has proven elusive.

Here we have developed a novel double graphene liquid cell (DGLC) to study the motion of individual, solvated metal atoms on an atomically thin membrane, in this case Pt adatoms on $MoS_2$. This system is a model single atom catalyst that has attracted significant attention due to its high activity, stability, and low cost as a catalyst in industrial applications[27,28]. Specifically, $MoS_2$ is utilised in hydrodesulphurization processes[29], and both single atom Pt[30] and $MoS_2$[31] have been studied individually for catalysing the hydrogen evolution reaction. A promising route to enhancing the catalytic properties of $MoS_2$ is by decorating the surface with noble metal nanoparticles and adatoms[2,32-34]. The interactions between the metal adatoms and $MoS_2$ are key to harnessing catalytic potential of these systems, and so have been intensively characterised by a variety of techniques: x-ray[35] and IR[36] spectroscopy, scanning tunnelling microscopy (STM)[37], and indeed *ex situ* STEM[28,38] under high vacuum conditions. Liquid cell imaging of the formation of Au nanoparticles on $MoS_2$ has been reported[39], but the resolution was limited to several nm. More recently a $MoS_2$ liquid cell (based on the original GLC design[23]) demonstrated the formation of aligned Pt nanocrystals on the $MoS_2$ surface, but lacked the resolution required to observe the atomic lattice sites or track dynamic motion of single metal atoms[40].

The novel DGLC design is illustrated in figure 1a. The structure consists of two lithographically patterned hBN spacer layers, each tens of nm thick, with a monolayer molybdenum disulphide ($MoS_2$) crystal sandwiched between them. Voids are pre-patterned in both hBN spacers using e-beam lithography and subsequent reactive ion etching. Liquid samples are then trapped inside the voids using few-layer graphene (FLG) on the top and bottom of the stack (see methods for the nanofabrication details). The atomically flat hBN crystal forms a hermetic seal with graphene and $MoS_2$; thus preventing leakage or liquid transfer between individual cells[26] and also preventing complete loss of liquid even if the membrane is damaged locally. Consequently, this DGLC design is highly robust, remaining intact under continuous STEM imaging for an electron flux of ~ 2.8 x $10^6$ e $s^{-1}$ $nm^{-2}$ (for more than 10 minutes at 200 kV). Interestingly, this design opens a unique future avenue to instigate mixing of separate liquid samples directly under the beam *via* directed ablation of the $MoS_2$ layer[41].

The HAADF STEM image in figure 1b shows a typical experimental sample (a larger field of view is shown in supplementary figure S3). The double liquid cell structure (shaded in yellow) can be found where the voids in both hBN layers overlap and the $MoS_2$ layer is thus suspended between two liquid pockets. Supplementary video 1 shows an image series of a double cell area at increasing magnification. An electron diffraction pattern from these double cell areas contains three sets of diffraction spots, which can be associated with the $MoS_2$ sheet and the two graphene windows (see supplementary figure S4). The presence of liquid within the voids can be verified by electron energy loss spectrum imaging (see supplementary figure S7) where the oxygen K-edge characteristic of water maps onto the double cell area. The samples were supported by a custom $SiN_x$ TEM support grid with large circular holes[42], outlined in green in figure 1b.

The DGLC is loaded with atomically-dispersed single atom Pt catalysts by filling the upper liquid cell pockets with an aqueous solution of platinum salt (10 mM $H_2PtCl_6$), while the lower part has been filled with deionised water. The Pt can be seen as bright spots in the HAADF STEM (Z-contrast) images due to the relatively high atomic number of Pt compared to other materials in the cell (verified as Pt by local elemental mapping, figure S6). Whether the Pt is adsorbed on the top graphene, the encapsulated MoS$_2$ monolayer, or the bottom graphene layer can be inferred using the finite focal depth (~10 nm) of the STEM probe (figure 1c - e), as shown in supplementary videos 2 - 3. A high density of individual atomic Pt species and nanometre-scale Pt nanocrystals are visible both on the top graphene window and on the submerged MoS$_2$ membrane (figures 1c and 1d respectively), while a few Pt atoms are also found on the bottom graphene window (top right region in figure 1e). Relative defocus values indicate a thickness of 42 nm for the upper liquid layer and 28 nm for the bottom layer; in good agreement with atomic force microscopy measurements of the hBN spacer layers acquired as part of the cell fabrication (32 nm and 30 nm respectively).

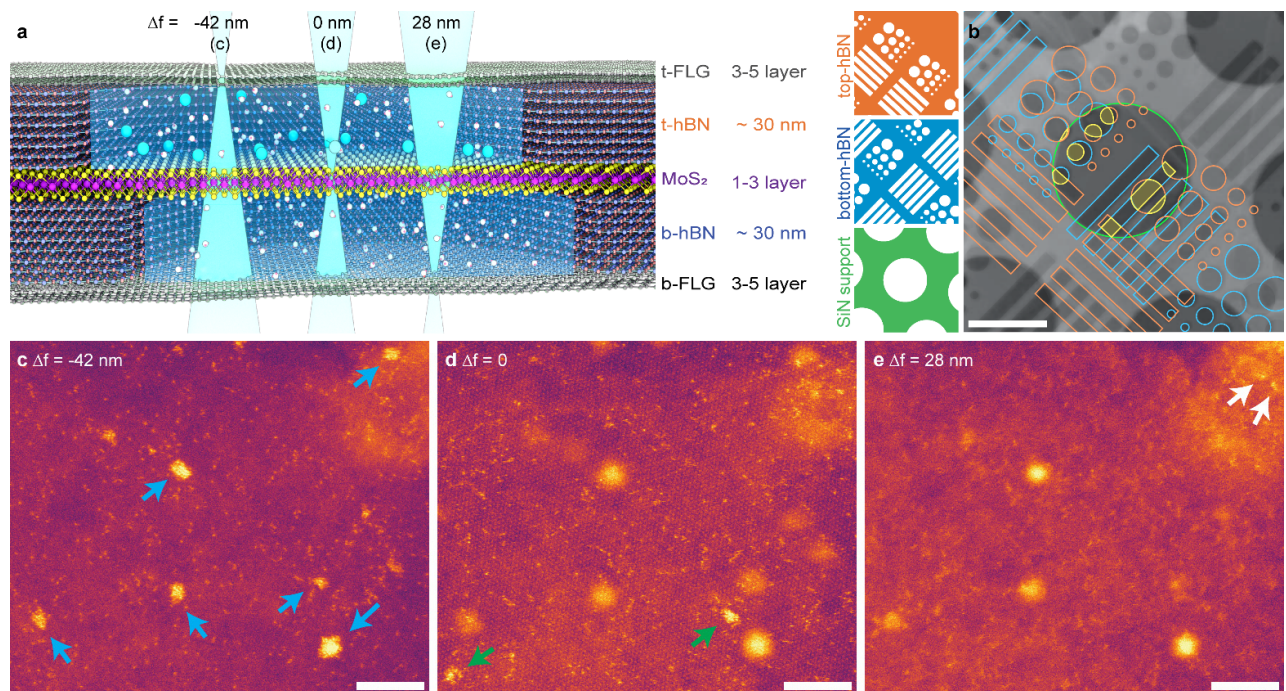

*Figure 1 Design of the double liquid cell. (a) Illustrated cross-section of a double liquid cell structure with labelled components. (b) A low magnification HAADF STEM image of the liquid cell arrays alongside colour-coded layer schematics, showing arrangements of the pockets in both hBN spacer layers and the SiN$_x$ support grid. (c-e) HAADF STEM images of one area of the double liquid cell with the probe focussed at the upper FLG window (Δf = -42 nm), the MoS$_2$ membrane (Δf = 0 nm), and the lower FLG membrane (Δf = 28 nm), as indicated by the probe convergence in (a). Images were recorded in the order presented (c-e). Arrows highlight selected parts of the image that are in focus at each defocus value. Scale bars (b) 2 μm, (c-e) 5 nm.*

The DGLC system allows investigation of both local coordination environment and the stability of preferred catalytic sites for single atom species in liquid. Let us first focus on analysis of the single atom lattice coordination. Figure 2 shows an image from a video sequence focused on the MoS$_2$ membrane (supplementary video 4), where the hexagonal lattice of the MoS$_2$ monolayer is decorated with brighter dots corresponding to single Pt species. The single atom catalysts have an approximately constant average areal density of ~ 0.6 nm$^{-2}$ (supplementary figure S13a). Image processing was employed to identify the lattice locations of the Pt on the supporting MoS$_2$ crystal: Pt positions were identified from Fourier filtered images, after which a template matching algorithm was used to reinforce local self-similarity and remove spatially uncorrelated noise, enabling identification of the Mo lattice sites. The two coordinate sets were then compared to find the Pt atom positions relative to the MoS$_2$ lattice. Full details of the image processing steps are presented in supplementary information Section 4.

This analysis shows that the Pt atoms prefer to occupy one of the three high symmetry MoS$_2$ lattice sites shown in figure 2b. These are referred to here as: the 'Mo site' where the Pt adatom is directly above the Mo lattice site; the 'S-site' where the Pt adatom is directly over the S-lattice site; and the hexagonal centre or 'HC site', where the Pt adatom is located above the hexagonal lattice site, equidistant from three S and three Mo sites. Figure 2c - d shows the experimental raw and averaged images demonstrating an excellent qualitative match to image simulations shown in figure 2e.

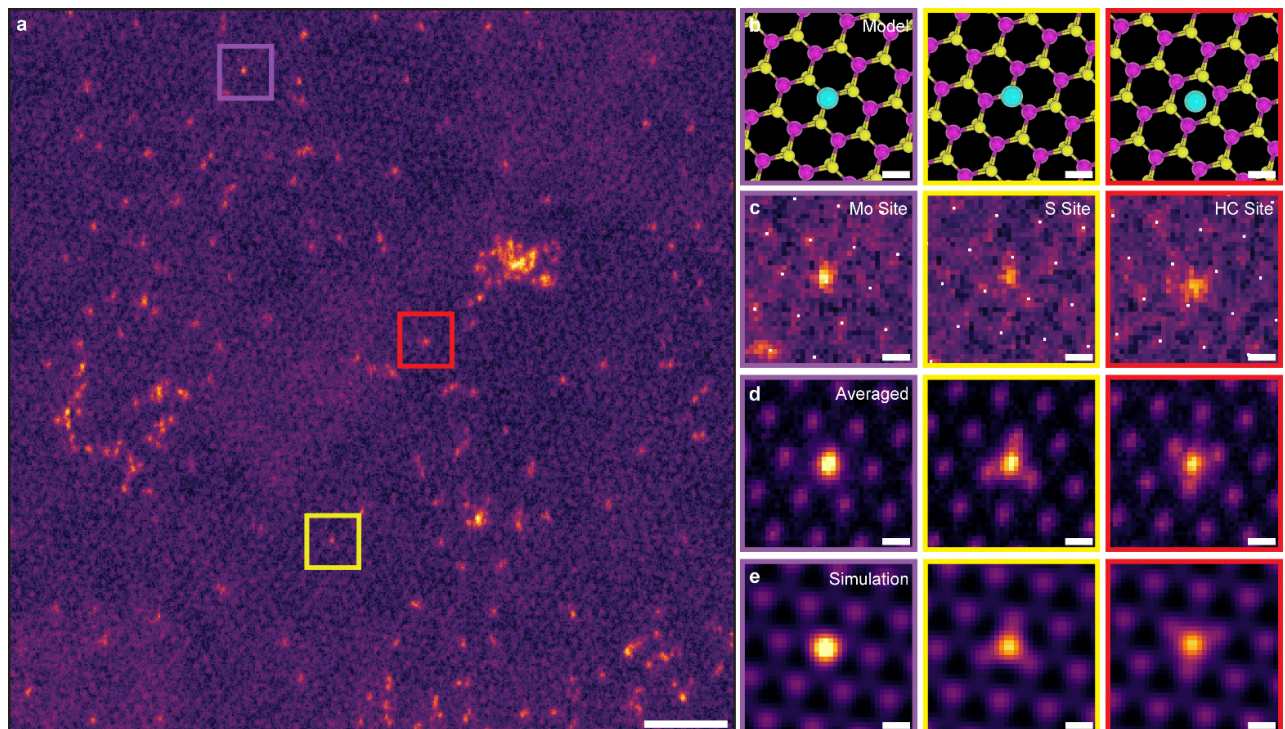

*Figure 2 Local lattice coordination of single Pt atoms with respect to underlying MoS$_2$ in an aqueous liquid environment.* (a) A representative HAADF STEM image from supplementary video 4. The probe is focussed on the submerged MoS$_2$ layer so that the Mo lattice and Pt adatoms are clearly visible (Δf = 0 nm). (b) Atomic models of a Pt adatom at a Mo site (outlined in a purple border), S site (yellow border), and hexagonal centre (HC) site (red border) on the MoS$_2$ flake. (c) Experimental image patches identified as showing atoms at the three lattice sites indicated in (a). Additional exemplar image patches are exhibited in supplementary figure S12. (d) Averaged images obtained by summation of 50 similar image patches. (e) Multi-slice image simulations of the model structures shown in (b). Scale bars (a) 2 nm (b-e) 200 pm.

To quantify the Pt atom coordination in liquid, we calculate a spatially resolved histogram showing the location of the adatoms relative to the underlying MoS$_2$ substrate, obtained from analysis of >70,000 Pt atoms (figure 3a-e). We also examine the distribution of Pt adatoms in a vacuum environment, with complementary histograms of preferred Pt lattice sites without liquid shown in figure 3f-j. This vacuum data was achieved both i) by deliberately puncturing the graphene windows of various cells, then allowing the liquid to escape by leaving the sample in the TEM vacuum overnight and ii) by preparing an *ex situ* Pt on MoS$_2$ sample for comparison. In vacuum, the experimental data shows that Pt atoms have a clear preference to sit above the S sites in the MoS$_2$ lattice, despite the fact that theoretical calculations indicate that the preferred position for Pt is above the Mo site in pristine MoS$_2$[43,44], with metastable states above S and HC sites. The strong preference for the S site in the experimental data is consistent with previous *ex situ* analysis and has been attributed to the presence of S vacancies[28] which present more energetically favourable positions with our density functional theory (DFT) calculations predicting a binding energy of 6.1 eV compared to 3.5 and 3.1 eV for Pt above Mo and S sites (see SI section 8). Indeed, by overlaying all occupied lattice positions in an image series, we see that the positions of Pt atoms are more evenly distributed in the liquid cell while the vacuum data is concentrated in specific locations (Figure S15). Such clustering is likely also associated with Pt pinning on less mobile S vacancies, which are expected to have comparatively low mobility compared to that of Pt adatoms on a perfect MoS$_2$ lattice (diffusion barrier of 0.8 - 2 eV for S vacancy vs ~ 0.5 eV for Pt adatom)[45,46]. Importantly, we observed the same relative adatom distribution on the freestanding MoS$_2$ membrane as we did for the dehydrated liquid cells, suggesting the distribution is relatively insensitive to the local chemistry in the absence of liquid.

Figure 3a-e reveals markedly different behaviour when comparing the vacuum results to the Pt lattice site distribution measured in the liquid cell, where we observe that both S and Mo sites display similar occupancies (figure 3a). This can be attributed to the substitution of S vacancies by oxygen from the water making them less attractive sites for Pt[47]. Indeed, our DFT calculations show that the binding energy for S vacancies is greater for O atoms (9.1 eV) than Pt atoms (6.1 eV) suggesting that where oxygen is available it will preferentially substitute into S vacancy sites (both the native S vacancies and those induced by the electron beam). Once oxygen substitutes a S vacancy the site becomes relatively unfavourable for a Pt adatom to locate (see figure S21b). Further details of the DFT calculations are provided in the supplementary information (SI section 8). This is consistent with our experimental observation that the Pt adatom resting behaviour is not dominated by strong S vacancy site interactions and more closely resembles the distribution predicted by DFT for pristine MoS$_2$. Nevertheless, we recognise that a full understanding of this experimental behaviour would benefit from detailed modelling of the possible different Pt species as well as the possible differences in the local binding energies due to the presence of a Pt hydration shell, although this is a huge computational undertaking. In both liquid and vacuum environments, we did not observe a dependency of the relative Pt distribution with time (electron dose), or with electron flux (over a range 0.7 – 2.8 x10$^6$ e nm$^{-2}$ s$^{-1}$), suggesting the behaviour we observe is relatively robust to electron beam induced changes in local environment (beam current details are provided in figure 3c-e and h-j). Furthermore, analysis of the Pt pair distribution function indicates the Pt adatoms prefer to sit next to each other which suggests incomplete hydration with no evidence to suggest long range coulombic repulsion (Figure S13e).

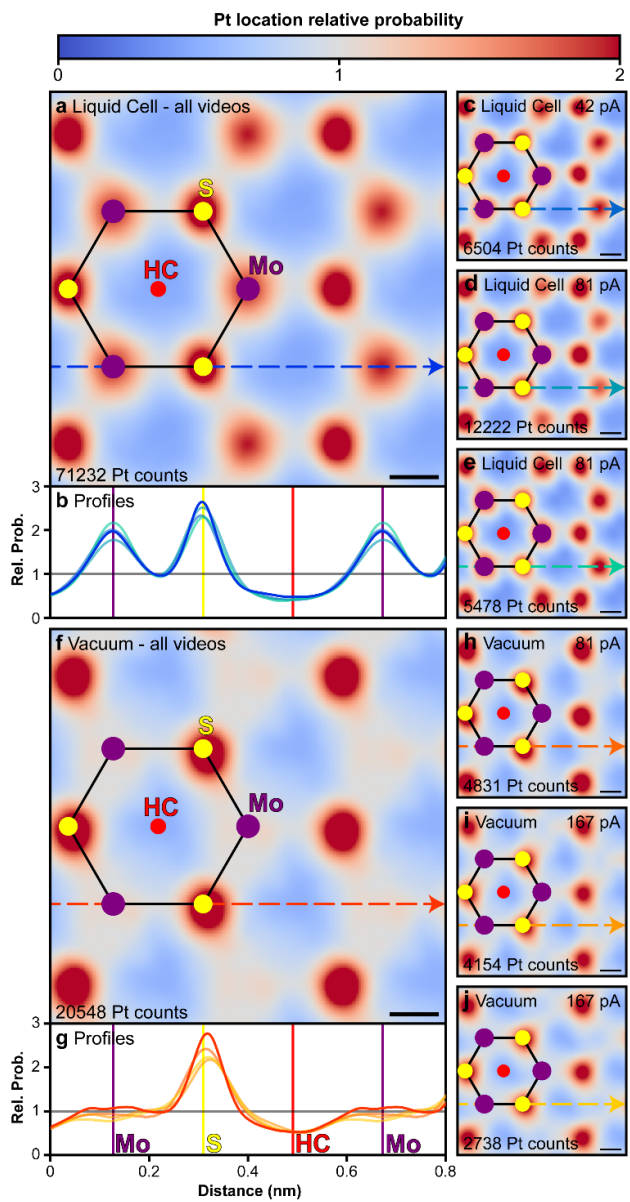

*Figure 3 Adatom preferred atomic lattice sites in liquid and vacuum.* *Spatially resolved histograms of Pt adatom positions relative to the MoS$_2$ lattice from (a) videos captured in liquid cell and (f) videos captured in vacuum environment. Mo, S, and HC positions are shown as an overlay by purple, yellow, and red circles respectively. Histograms from individual experiments for various electron dose rates and different samples are shown in (c-e) for liquid and (h-j) for vacuum environments where the relevant electron probe current is given top left. Videos in (h-i) are acquired from dehydrated liquid cells and (j) from a freestanding Pt-decorated MoS$_2$ membrane. The histograms are averaged to one MoS$_2$ unit cell and tiled in the figure for clarity, so all Mo and S points are equivalent. Probability density line profiles are shown in (b) for liquid and (g) for vacuum environments, extracted along positions indicated by the arrows of corresponding colour from each histogram.*

We now turn to discussion of the dynamic motion of the Pt atoms in liquid, compared to that observed in vacuum, by linking the atomic positions in adjacent frames using a minimum displacement approach (full details in supplementary information). The vast majority of single Pt atoms are observed to be mobile, as demonstrated in figure 4a - b, which shows the trajectories of several representative Pt atoms over a sustained period (the full image series for this dataset comprises supplementary videos 6 and 7). Statistical analysis of these atomic dynamics for both liquid cell and for vacuum conditions, as well as a range of electron fluxes, yields histograms of radially averaged, frame-to-frame, atomic displacements that are plotted in figure 4c. Both distributions are peaked at Pt jump distances of ~ 0.05 nm, which is likely related to motion around a chosen lattice site. Note that this motion is not caused by tracking uncertainty or the background movement of the underlying MoS$_2$, which is tracked as the displacement of Mo sites after correcting for specimen drift, tilt, and rotation, and is found to have much smaller peak values (below 0.02 nm for both liquid and vacuum conditions, inset figure 4c). Note that for both systems, although vacancies in the underlying MoS$_2$ lattice can influence the behaviour we observe, the high mobility of Pt atoms is more consistent with their being predominantly surface adatoms: located on, rather than substituted in, the MoS$_2$ lattice. This is supported by our DFT calculations where the binding energy of a Pt substitution to a S vacancy is greater than the energy that can theoretically be provided by a 200 keV electron (see SI section 8). The only exception was a small number of fully immobilized Pt atoms. These were located at Mo sites suggesting they are Pt substituted into a Mo vacancy (see supplementary figure S5). In addition, we note that in the liquid cell adatoms occasionally appear and disappear spontaneously during imaging, likely due to a continuous cycle of Pt deposition onto the MoS$_2$ surface from the salt solution and dissolution of Pt adatoms back into solution.

Comparing liquid and vacuum data sets, the distribution of atomic displacements in the liquid cell data is noticeably broader, demonstrating more frequent occurrence of larger (0.1 - 0.5 nm) frame-to-frame displacements between different sites. Considering the mean square displacements (MSDs), it is evident that larger displacements take place in the liquid cell relative to the sample imaged in vacuum, yielding surface diffusivities, $D$, for Pt of >0.25 and <0.2 nm$^2$ s$^{-1}$, for liquid and vacuum imaging respectively. The MSD measured for the Mo atoms themselves is two orders of magnitude smaller, after drift correction, so MoS$_2$ substrate movement is considered negligible. In vacuum conditions, the energy barrier for Pt adatom diffusion between adjacent Mo sites on a MoS$_2$ lattice is calculated to be 0.5 – 0.82 eV[28,45] (SI section 8). The diffusion barrier of charged Pt surface ions (both reduced and oxidised) decreases by ~0.2 eV (SI section 8), which may contribute to the larger $D$ of the liquid cell, although distinguishing the local charge state for individual atoms is not generally possible even for *ex situ* TEM studies. An analysis of Pt-Pt adatom distances for liquid and vacuum conditions does not reveal preferred separation distance, which might indicate a hydration layer or short-range Coulomb forces. Nonetheless, this may be due to the dominance of the metal-support interactions, the investigation of which in fully aqueous conditions would require a huge computational effort.

Combining the lattice site identification and dynamic atom tracking allows additional analysis of the statistics and directionality of site hopping (figure S19c). The nearly perfect symmetry seen in this data for all crystallographically equivalent lattice directions suggests that the electron beam scanning does not have a strong effect on hopping directionality. For first nearest neighbour hopping, the Pt atoms were more likely to hop from a S (Mo) site to a neighbouring Mo (S) site than into an equidistant HC site. This was seen in both liquid and vacuum, although the effect was more pronounced in the liquid cell consistent with the histogram lattice occupancy. In addition, unlike the vacuum system, Pt atoms atop S-sites are less likely to be stuck for multiple frames (figure S19a,b), which supports our hypothesis that the Pt's bonding to S vacancies is inhibited in liquid compared to vacuum conditions.

It is important to note that, similar to many liquid environments relevant to catalysis and nucleation, this aqueous experimental system is highly complex and contains a wide range of chemical species. The calculated diffusion barriers are far below the typical energy which can be transferred to a Pt adatom under 200 kV illumination (2.69 eV calculated according to ref.[48]) indicating that, similar to all TEM investigations of atomic surface dynamics, the energy transfer from the electron beam is likely to be the largest trigger of the atomic movement we observe. Furthermore, the electron beam is known to directly affect the environmental conditions including modifying pH and the relative concentrations of metallic species in the liquid and adsorbed on the substrate[21]. To better understand the sensitivity of the Pt adatom resting sites and dynamic motion to the electron beam we have tested a range of electron flux conditions produced by changing the electron probe current. As illustrated in Figure 3 and Figure 4 these results show that the measured differences between atom behaviour in liquid and vacuum conditions are largely independent of the electron flux used. Compared to liquid cell studies using traditional TEM imaging, STEM generally requires a higher local electron flux but the localised electron probe also provides more opportunity for beam induced species to diffuse away from the active imaging area, providing opportunities restoring the pre-irradiation conditions. Additionally, the use of ultra-thin and highly conductive graphene windows will greatly reduce irradiation induced artefacts compared to traditional silicon nitride liquid cells[22], which is likely contributing to the insensitivity of the results to the precise imaging conditions.

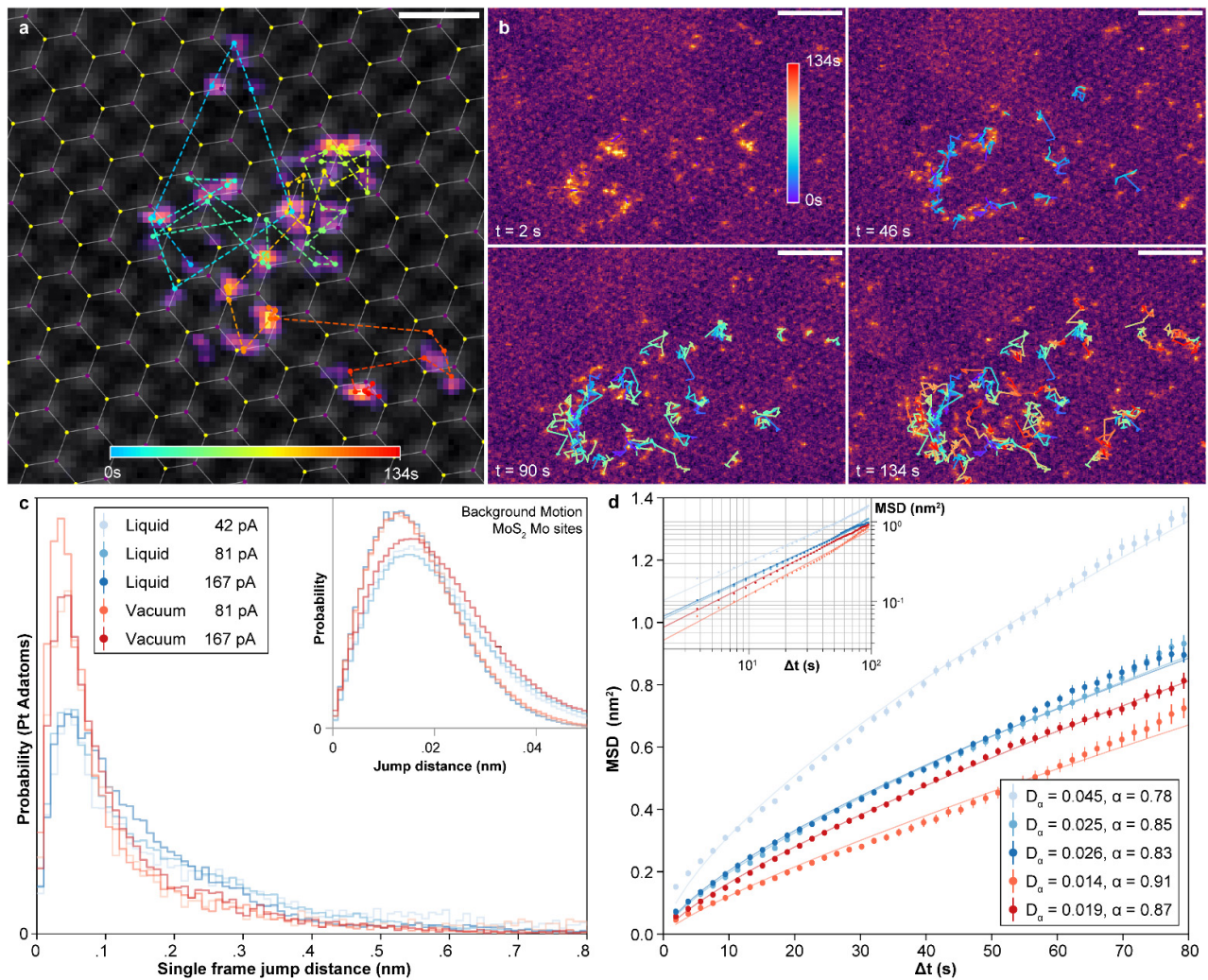

*Figure 4 Single atom tracking using nearest neighbour linking.* (a) A single Pt adatom trajectory from a 134s video in the graphene double liquid cell (supplementary video 8). (b) Trajectories of an ensemble of Pt atoms in liquid cell (from supplementary video 4), coloured according to the elapsed time (see supplementary figure S14 for wider view). (c) Histogram showing relative frame-to-frame jump distances for Pt atoms averaged over multiple videos in liquid (blue lines) and vacuum environments (red lines), captured with 3 beam currents, and with all distributions peaked at ~ 0.05nm. Inset compares the background movement of the underlying $MoS_2$ (tracked from the Mo sites) which has much smaller peak values of under 0.02 nm. (d) The averaged mean squared displacement as a function of the lag time $\Delta t$ plotted for Pt adatoms in both environments (inset shows the same dataset plotted with data on logarithmic axes). The bars represent standard error. A least-squares fit to a power law function ($MSD(\Delta t) = D_\alpha \Delta t^\alpha$) is shown overlaid and the corresponding fit parameters ($D_\alpha$, $\alpha$) are listed in the lower right. Scale bars (a) 0.5 nm (b) 2 nm.

In summary, we have demonstrated a double graphene liquid cell concept and applied this to investigate single Pt adatom lattice coordination and dynamic motion on a submerged monolayer MoS$_2$ membrane. The combination of a novel liquid cell platform and advanced image analysis techniques have enabled the first demonstration of single atom tracking in liquid. The measurements made in this work, based on the individual dynamics of ~70,000 adatom resting lattice sites and hopping behaviour relative to the MoS$_2$ lattice, have yield new, quantitative insights into the behaviour of atoms regarding their chemical and physical interactions with a substrate crystal under two extreme conditions: full hydration and high vacuum. Comparing our liquid cell results to *ex situ* vacuum conditions, we find a weaker dependency on S vacancy sites for the Pt location, with both Mo and S sites strongly preferred resting sites in liquid. We also measure higher diffusivities and a broader range of atomic jump distances for the liquid cell compared to vacuum data. The new experimental platform demonstrated here provides a pathway to previously inaccessible structural information regarding single atom motion in the liquid phase, opening an avenue to explore many diverse systems across the physical sciences.

# References

1. Jones, J. *et al.* Thermally stable single-atom platinum-on-ceria catalysts via atom trapping. *Science* **353**, 150-154 (2016).
2. Li, H. *et al.* Synergetic interaction between neighbouring platinum monomers in CO$_2$ hydrogenation. *Nat. Nanotechnol.* **13**, 411-417 (2018).
3. Daelman, N., Capdevila-Cortada, M. & López, N. Dynamic charge and oxidation state of Pt/CeO$_2$ single-atom catalysts. *Nat. Mater.* **18**, 1215-1221 (2019).
4. Kyriakou, G. *et al.* Isolated Metal Atom Geometries as a Strategy for Selective Heterogeneous Hydrogenations. *Science* **335**, 1209-1212 (2012).
5. Nie, L. *et al.* Activation of surface lattice oxygen in single-atom Pt/CeO$_2$ for low-temperature CO oxidation. *Science* **358**, 1419 (2017).
6. Wang, A., Li, J. & Zhang, T. Heterogeneous single-atom catalysis. *Nat. Rev. Chem.* **2**, 65-81 (2018).
7. Sharapa, D. I., Doronkin, D. E., Studt, F., Grunwaldt, J.-D. & Behrens, S. Moving Frontiers in Transition Metal Catalysis: Synthesis, Characterization and Modeling. *Adv. Mater.* **31**, 1807381 (2019).
8. Lee, J., Yang, J., Kwon, S. G. & Hyeon, T. Nonclassical nucleation and growth of inorganic nanoparticles. *Nat. Rev. Mater.* **1**, 16034 (2016).
9. Qiao, B. *et al.* Single-atom catalysis of CO oxidation using Pt$_1$/FeO$_x$. *Nat. Chem.* **3**, 634-641 (2011).
10. Nellist, P. D. & Pennycook, S. J. Direct Imaging of the Atomic Configuration of Ultradispersed Catalysts. *Science* **274**, 413 (1996).
11. de Jonge, N. & Ross, F. M. Electron microscopy of specimens in liquid. *Nat. Nanotechnol.* **6**, 695-704 (2011).
12. Tao, F. & Salmeron, M. In Situ Studies of Chemistry and Structure of Materials in Reactive Environments. *Science* **331**, 171 (2011).
13. Vendelbo, S. B. *et al.* Visualization of oscillatory behaviour of Pt nanoparticles catalysing CO oxidation. *Nat. Mater.* **13**, 884-890 (2014).
14. Boyes, E. D., LaGrow, A. P., Ward, M. R., Mitchell, R. W. & Gai, P. L. Single Atom Dynamics in Chemical Reactions. *Acc. Chem. Res.* **53**, 390-399 (2020).
15. DeRita, L. *et al.* Structural evolution of atomically dispersed Pt catalysts dictates reactivity. *Nat. Mater.* **18**, 746-751 (2019).
16. Ross, F. Opportunities and challenges in liquid cell electron microscopy. *Science* **350** (2015).
17. de Jonge, N., Houben, L., Dunin-Borkowski, R. E. & Ross, F. M. Resolution and aberration correction in liquid cell transmission electron microscopy. *Nat. Rev. Mater.* **4**, 61-78 (2019).
18. Park, J. *et al.* 3D structure of individual nanocrystals in solution by electron microscopy. *Science*

**349**, 290 (2015).

19   Lee, C., Wei, X. D., Kysar, J. W. & Hone, J. Measurement of the elastic properties and intrinsic strength of monolayer graphene. *Science* **321**, 385-388 (2008).

20   Sun, P. Z. *et al.* Limits on gas impermeability of graphene. *Nature* **579**, 229-232 (2020).

21   Woehl, T. J. & Abellan, P. Defining the radiation chemistry during liquid cell electron microscopy to enable visualization of nanomaterial growth and degradation dynamics. *J. Microsc.* **265**, 135-147 (2017).

22   Cho, H. *et al.* The Use of Graphene and Its Derivatives for Liquid-Phase Transmission Electron Microscopy of Radiation-Sensitive Specimens. *Nano Lett.* **17**, 414-420 (2017).

23   Yuk, J. M. *et al.* High-Resolution EM of Colloidal Nanocrystal Growth Using Graphene Liquid Cells. *Science* **336**, 61-64 (2012).

24   Textor, M. & de Jonge, N. Strategies for Preparing Graphene Liquid Cells for Transmission Electron Microscopy. *Nano Lett.* **18**, 3313 (2018).

25   Rasool, H., Dunn, G., Fathalizadeh, A. & Zettl, A. Graphene-sealed Si/SiN cavities for high-resolution in situ electron microscopy of nano-confined solutions. *Phys. Status Solidi B* **253**, 2351-2354 (2016).

26   Kelly, D. J. *et al.* Nanometer Resolution Elemental Mapping in Graphene-Based TEM Liquid Cells. *Nano Lett.* **18**, 1168 (2018).

27   Miremadi, B. K. & Morrison, S. R. Exfoliated MoS2 for stabilization and activation of Pt oxidation catalysts. *Journal of Catalysis* **131**, 127-132 (1991).

28   Li, H. *et al.* Atomic Structure and Dynamics of Single Platinum Atom Interactions with Monolayer $MoS_2$. *ACS Nano* **11**, 3392-3403 (2017).

29   Mom, R. V., Louwen, J. N., Frenken, J. W. M. & Groot, I. M. N. In situ observations of an active $MoS_2$ model hydrodesulfurization catalyst. *Nat. Commun.* **10**, 2546 (2019).

30   Cheng, N. *et al.* Platinum single-atom and cluster catalysis of the hydrogen evolution reaction. *Nat. Commun.* **7**, 13638 (2016).

31   Jaramillo, T. F. *et al.* Identification of Active Edge Sites for Electrochemical $H_2$ Evolution from $MoS_2$ Nanocatalysts. *Science* **317**, 100-102 (2007).

32   Deng, D. *et al.* Catalysis with two-dimensional materials and their heterostructures. *Nat. Nanotechnol.* **11**, 218 (2016).

33   Huang, X. *et al.* Solution-phase epitaxial growth of noble metal nanostructures on dispersible single-layer molybdenum disulfide nanosheets. *Nat. Commun.* **4**, 1444 (2013).

34   Li, S., Lee, J. K., Zhou, S., Pasta, M. & Warner, J. H. Synthesis of Surface Grown Pt Nanoparticles on Edge-Enriched $MoS_2$ Porous Thin Films for Enhancing Electrochemical Performance. *Chem. Mater.* **31**, 387-397 (2019).

35   Choi, C. H. *et al.* Tuning selectivity of electrochemical reactions by atomically dispersed platinum catalyst. *Nat. Commun.* **7**, 10922 (2016).

36   Ding, K. *et al.* Identification of active sites in CO oxidation and water-gas shift over supported Pt catalysts. *Science* **350**, 189 (2015).

37   Liu, J. *et al.* Tackling CO Poisoning with Single-Atom Alloy Catalysts. *J. Am. Chem. Soc.* **138**, 6396-6399 (2016).

38   Lin, Y.-C. *et al.* Properties of Individual Dopant Atoms in Single-Layer $MoS_2$: Atomic Structure, Migration, and Enhanced Reactivity. *Adv. Mater.* **26**, 2857-2861 (2014).

39   Song, B. *et al.* In situ study of nucleation and growth dynamics of Au nanoparticles on $MoS_2$ nanoflakes. *Nanoscale* **10**, 15809-15818 (2018).

40   Yang, J. *et al.* $MoS_2$ Liquid Cell Electron Microscopy Through Clean and Fast Polymer-Free $MoS_2$ Transfer. *Nano Lett.* **19**, 1788-1795 (2019).

41   Kelly, D. J. *et al.* In Situ TEM Imaging of Solution-Phase Chemical Reactions Using 2D-

Heterostructure Mixing Cells. *Adv. Mater.* **n/a**, 2100668 (2021).

42  Hamer, M. J. *et al.* Atomic Resolution Imaging of CrBr3 Using Adhesion-Enhanced Grids. *Nano Lett.* **20**, 6582-6589 (2020).

43  Chen, D., Zhang, X., Tang, J., Cui, H. & Li, Y. Noble metal (Pt or Au)-doped monolayer $MoS_2$ as a promising adsorbent and gas-sensing material to $SO_2$, $SOF_2$ and $SO_2F_2$: a DFT study. *Appl. Phys. A* **124**, 194 (2018).

44  Chang, J., Larentis, S., Tutuc, E., Register, L. F. & Banerjee, S. K. Atomistic simulation of the electronic states of adatoms in monolayer $MoS_2$. *Appl. Phys. Lett.* **104**, 141603 (2014).

45  Wu, P., Yin, N., Li, P., Cheng, W. & Huang, M. The adsorption and diffusion behavior of noble metal adatoms (Pd, Pt, Cu, Ag and Au) on a $MoS_2$ monolayer: a first-principles study. *Phys. Chem. Chem. Phys.* **19**, 20713-20722 (2017).

46  Komsa, H.-P., Kurasch, S., Lehtinen, O., Kaiser, U. & Krasheninnikov, A. V. From point to extended defects in two-dimensional $MoS_2$: Evolution of atomic structure under electron irradiation. *Phys. Rev. B* **88**, 035301 (2013).

47  Lu, J. *et al.* Atomic Healing of Defects in Transition Metal Dichalcogenides. *Nano Lett.* **15**, 3524-3532 (2015).

48  Garcia, A. *et al.* Analysis of electron beam damage of exfoliated $MoS_2$ sheets and quantitative HAADF-STEM imaging. *Ultramicroscopy* **146**, 33-38 (2014).

## Methods

**Solutions:** Hexachloroplatinic acid ($H_2PtCl_6$, 8 wt. % in water) was purchased from Sigma Aldrich UK, and was diluted down to a 10 mM solution in 3:1 deionised water: propan-2-ol.

**Liquid Cell Fabrication:** 2D materials (~ 30 nm thick hBN flakes, 2 - 3 nm thick FLG flakes, and monolayer $MoS_2$) were mechanically exfoliated from bulk crystals onto thermally oxidised silicon substrates. Synthetic hBN and natural graphite and $MoS_2$ were purchased from HQ Graphene (hqgraphene.com), NaturGraphit GmbH (graphit.de), and Manchester Nanomaterials (mos2crystals.com) respectively. The hBN flakes were patterned using electron beam lithography to define a resist etch mask, and subsequent reactive ion etching with $SF_6/O_2$. Flakes were then successively stacked, inserting liquid samples (10mM in 3:1 water: propan-2-ol in upper cell, 3:1 water: propan-2-ol in lower cell) between flakes immediately before bringing them into contact in order to trap the liquid sample in the defined pockets patterned in the hBN spacer layers. The stack was then transferred to a custom silicon nitride support film for STEM characterisation (supplementary figures S1 - 2). After imaging, specific liquid cells were perforated using prolonged exposure to a focused electron beam, and left to dehydrate in the TEM column overnight, enabling us to compare Pt diffusion for the same samples in liquid and vacuum environments.

**Vacuum sample fabrication:** $MoS_2$ flakes were exfoliated in the same way as for the liquid cell, from the identical $MoS_2$ crystal used for the double liquid cell membrane. A monolayer flake was transferred onto a custom silicon nitride support film and annealed in a reducing atmosphere to remove transfer contamination (200C in 5% H in Ar). A 20 mM Hexachloroplatinic acid aqueous solution was drop cast onto the grid, which was then baked on a hot plate in argon atmosphere to evaporate any remaining liquid.

**STEM Imaging:** High resolution imaging and spectroscopy was carried out using a FEI Titan G2 80 - 200 S/TEM ChemiSTEM. The microscope was operated in STEM mode with a 200 kV accelerating voltage and a probe current of 160 pA, a 21 mrad convergence angle, and a 48 mrad HAADF collection angle. Probe aberrations up to $3^{rd}$ order were corrected to better than a $\pi/4$ phase shift at 20 mrad. We find that the cell is stable under continuous STEM imaging with an electron flux of ~ $2.8 \times 10^6$ $e^-$ $s^{-1}$ $nm^{-2}$ (for more than 10 minutes at 200 kV).

**STEM Image simulation:** Multislice image simulation of the $MoS_2$ / Pt adatom structure in vacuum was carried out using QSTEM[49]. Image simulation parameters were chosen to simulate the experimental imaging conditions used in the liquid cell, with a virtual electron source size of 1.5 Å. Thermal effects were simulated by averaging the output over 25 random phonon configurations, using default Debye-Waller (DW) factors calculated by QSTEM ($Mo_{DW} = 0.131$, $S_{DW} = 0.393$, $Pt_{DW} = 0.065$). The simulations did not include the atomic structures of the encapsulating graphene or liquid layers. Previous work has shown that exclusion of the graphene has little qualitative effect on the relative peak intensities of heavy atom species[50].

**Pt Adatom Tracking:** Individual HAADF STEM images from the videos were processed using a template matching[51]/similarity search[52] and reconstruction algorithm to highlight the Mo sites. A comparison between the original and reconstructed image is shown in supplementary figure S11. Mo lattice sites in each frame were found in this processed image and linked by nearest neighbour frame-to-frame. Pt adatom locations were separately identified from a Fourier space filtered version of the original videos, after drift correction (pixelwise, translation only) using the mean Mo lattice displacement. Pt adatom trajectories were generated by linking nearest neighbour Pt sites in adjacent frames. Linking of identified particles was performed using an implementation of the Crocker-Grier[53] algorithm contained in the trackpy package[54]. The matrix transformation between the Mo sites in each frame and their time averaged positions was then found, and used to apply an affine transformation (translation, scale, rotation, shear) to remove the effects of time varying scan distortions from the identified atomic trajectories. For the liquid cell, 15452 positions of Pt atoms are compared, with 3670 images from the vacuum sample image series to generate the statistics shown in figures 3 and 4, and S13 - S19 in the supplementary information. Atom positions in the vicinity of dense clusters, such as that in supplementary video 4 (as shown in figure S16) were not considered.

See supplementary information for a full description. An example workflow is shown in figure S10.

**DFT Calculations:** Spin-polarised density functional theory calculations with the optB86b-vdW functional[55] were performed in conjunction with projector augmented wave pseudopotentials within a 500 eV energy cutoff and a (3 x 3 x 1) k-point mesh as well as dipole corrections using the Vienna Ab-initio Simulation Package (VASP)[56] For realistic modelling of defects and diffusion processes, we used a (5 × 5 × 1) $MoS_2$ supercell slab with 16 Å vacuum. Every atom in the unit cell is fully relaxed in all calculations with the force threshold of 0.01 eV/Å for structure optimisation and of 0.02 eV/Å for transition pathway search.

## Data Availability

Additional and raw data related to this paper is available from the corresponding authors upon reasonable request.

## Code Availability

The computer code used for the image processing is available from the corresponding authors upon reasonable request.

## Acknowledgements

The work was supported by EPSRC grants EP/M010619/1, EP/S021531/1 and EP/P009050/1 and EPSRC Doctoral Prize Fellowship, the European Research Council (ERC) under the European Union's Horizon 2020 research and innovation programme (Grant ERC-2016-STG-EvoluTEM-715502, the ERC Synergy Hetero2D project 319277, the European Graphene Flagship Project (696656), and the European Quantum Technology Flagship Project 2DSIPC (820378). D.J.H and D.J.K. acknowledge the EPSRC NoWNano programme for funding. R.G. acknowledges funding from the Royal Society. C.W.M. was supported by NRF Korea (NRF-2020R1A6A3A03039808), Swiss National Supercomputing Centre (s1052) and KISTI (KSC-2021-CRE-0129). C.S. acknowledges partial financial support from the *Alexander von Humboldt-Stiftung*. C.W.M., C.S. and A.M. acknowledge the support from ARCHER, for which access was obtained via the UKCP consortium and funded by EPSRC (EP/P022561/1).

## Author Contributions

R.G and S.J.H conceived the project and supervised the work. The liquid cell fabrication was performed by M.Z, N.C. and R.G. TEM imaging was performed by D.J.K, Y-C.Z and S.J.H and data analysis was performed by N.C with support from D.J.K. Image simulations were performed by D.G.H. C.W.M. performed the DFT calculations and C.W.M., C.S., and A.M. analysed the DFT data.

The authors declare no competing interests.

## Additional Information

Supplementary Information is available for this paper.

Correspondence and requests for materials should be addressed to S.J.H or R.G.

# Methods References

49  Koch, C. T. *Determination of core structure periodicity and point defect density along dislocations*, Arizona State University, (2002).

50  Hopkinson, D. G. *et al.* Formation and Healing of Defects in Atomically Thin GaSe and InSe. *ACS Nano* **13**, 5112-5123 (2019).

51  Zuo, J.-M. *et al.* Lattice and strain analysis of atomic resolution Z-contrast images based on template matching. *Ultramicroscopy* **136**, 50-60 (2014).

52  Somnath, S. *et al.* Feature extraction via similarity search: application to atom finding and denoising in electron and scanning probe microscopy imaging. *Advanced Structural and Chemical Imaging* **4**, 3 (2018).

53  Crocker, J. C. & Grier, D. G. Methods of Digital Video Microscopy for Colloidal Studies. *J. Colloid. Interf. Sci.* **179**, 298-310 (1996).

54  Allan, D. *et al.* soft-matter/trackpy: Trackpy v0.4.2. *Zenodo* (2019).

55  Klimeš, J., Bowler, D. R. & Michaelides, A. Chemical accuracy for the van der Waals density functional. *Journal of Physics: Condensed Matter* **22**, 022201 (2009).

56  Kresse, G. & Furthmüller, J. Efficient iterative schemes for ab initio total-energy calculations using a plane-wave basis set. *Phys. Rev. B* **54**, 11169-11186 (1996).